# A RESONANT CAVITY FOR SINGLE-SHOT EMITTANCE MEASUREMENT*

J. S. Kim[+], FARTECH, Inc., San Diego, CA, 92122
C. D. Nantista, D. H. Whittum, R. H. Miller, S. G. Tantawi, SLAC, CA 94309
A. W. Weidemann, University of Tennessee, Knoxville, TN 37996

*Abstract*

We present a non-invasive, resonant cavity based approach to beam emittance measurement of a shot-to-shot non-circular beam pulse of multi-bunches. In a resonant cavity, desired field components can be enhanced up to $Q_{L\lambda}/\pi$, where $Q_{L\lambda}$ is the loaded $Q$ of the resonance mode $\lambda$, when the cavity resonant mode matches with the beam operating frequency. In particular, a Quad-cavity, with its quadrupole mode at beam operating frequency, extracts the beam quad-moment exclusively, utilizing the symmetry of the cavity and some simple networks to suppress common modes. Six successive beam quadrupole moment measurements, performed at different betatron phases in a linear transport system, allow us to determine the beam emittance, i.e., the beam size and shape in the beam's phase space. One measurement alone provides the rms-beam size if the beam position is given, for instance, by nearby beam-position-monitors. This paper describes the basic design and analysis of a Quad-cavity beam monitoring system.

## 1 INTRODUCTION

RMS beam emittance is a key beam parameter, along with beam position, for accelerator operations[1]. Currently single-pulse emittance measurement is not available, and thus beam tune-up time takes from hours to days. A pulse-to-pulse size measurement based on striplines has been proposed previously, by Miller, *et al.*[2]. We have extended the idea of a resonant cavity[3] as a beamline instrument for beam emittance measurement. Advantages over a stripline configuration include: (1) stronger beam and desired cavity mode coupling, *i.e.*, larger [R/Q]; and (2) high signal-to-noise ratio by resonance of a cavity mode at the bunch frequency.

The resonant cavity approach for emittance measurement is to employ a resonant cavity monitor operated in the quadrupole mode. This provides a voltage phasor output proportional to the product of charge and the beam moment $<x_L^2 - y_L^2>$ in the linac coordinates, phased with the beam. A series of such monitors placed in a FODO lattice, and separated by adequate machine phase advance, permits one to deconvolve beam matching parameters, and rms emittance.

* Work supported by US Department of Energy under the grants DE-FG03-98ER82574, and DE-AC03-76SF00515.
+ kimjs@far-tech.com

The Quad-cavity is designed, by symmetry, to have a quadrupole mode that is exclusively excited by the beam's quadrupole moment, and thus, for a flat beam, is indicative of its size. In order for maximum coupling between the beam quadrupole moment $<x_L^2 - y_L^2>$ and the cavity quad-mode, the cavity must be rotated by 45 degrees along the linac axis as in Figure 1.

Parasitic common modes, i.e. modes other than the quad-modes, are minimized by: (1) separating the other resonance mode away from the quad-mode by at least 1 GHz; (2) maximizing the beam and quad-mode coupling by optimizing the cavity length along the beam pipe; (3) utilizing symmetry in simple network to cancel out parasitic modes, and (4) using for this network waveguide in which the lower modes are cutoff. The electric field can be reduced by many orders of magnitude at a frequency about 1 GHz away from the mode frequency. Further, the quad mode is the lowest mode of odd-odd symmetry with respect to the planes of x=0 and y=0, in the cavity coordinates. Thus, the monopole and dipoles can be superimposed away via proper hybrid T connections, as shown in Figure 1.

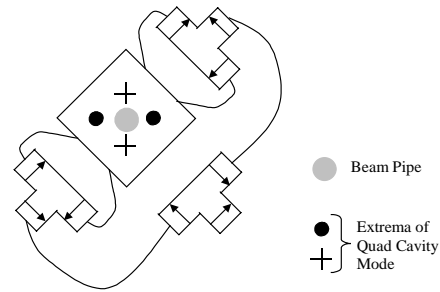

Figure 1. Cavity orientation in the beamline.

## 2 CIRCUIT MODEL ANALYSIS

The time-behavior of the voltage radiated from a beam-driven cavity mode may be described as a driven, damped, simple harmonic oscillator. With a drive term proportional to beam current $I_b$ and appropriate geometric factors and damping inversely proportional to the loaded quality factor of the mode $Q_{L\lambda}$,

$$\left(\frac{d^2}{dt^2} + \frac{\omega_\lambda}{Q_{L\lambda}}\frac{d}{dt} + \omega_\lambda^2\right)V_\lambda(t, r_\perp) \approx -2k_\lambda \frac{d}{dt}I_b X_\lambda(r_\perp),$$

where we abbreviated $X_\lambda = 1, \langle x \rangle, \langle xy \rangle$, and $k_\lambda = k_m, k'_d, k''_q$, $k'_d \equiv k_d/x$ for x-dipole, $k''_q \equiv k_q/(x^2 y^2)$ for quad. We assumed a perfect match on the output line. The solution of the circuit equation for the induced mode voltage driven by a single-bunch with a Gaussian distribution in z of rms-length $\sigma_t$, may be expressed as a damped sinusoidal signal when $\omega_\lambda \sigma_t \ll 1$. Further, for $Q_{L\lambda} \gg 1$:

$$V_\lambda(t_r) = 2 k_\lambda Q_\lambda X_\lambda \Re[\exp(-\Gamma(t-t_b))]$$

where $\Gamma \equiv \dfrac{\omega_\lambda}{2Q_{L\lambda}} - i\omega_\lambda$ and $\Re$ refers to the real part of the function. The amplitude of the sinusoidal voltage is proportional to $|V_\lambda|$. Multi-bunch responses can be obtained by summing over the single bunch responses at times delayed by the bunch interval $\tau$. For a train of N bunches, the amplitudes of the sinusoidal voltage of the $\ell$ th-bunch are,

$$\left| V_\lambda^N(t_\ell)/V_\lambda^N(t_0) \right| = \left| (1-\exp(-\Gamma\ell\tau))/(1-\exp(-\Gamma\tau)) \right|$$

where $t_\ell = t_0 + (\ell-1)\tau$ for $\ell = 1,2,\cdots,N$, and for $t > t_N$

$$\left| V_\lambda^N(t > t_N) \right| = \left| V_\lambda^N(t_N) \exp(-\Gamma(t-t_N)) \right|.$$

Figure 2 shows the amplitudes of voltages of 1000 bunches, filling every potential bucket, when the mode frequency is at perfect resonance with the bunch frequency (upper curve), at a frequencies mismatched by $\delta f = f_0/(2Q) = 7.6$ MHz with $Q = 750$ (middle curve), and by $2\,\delta f$ (bottom curve).

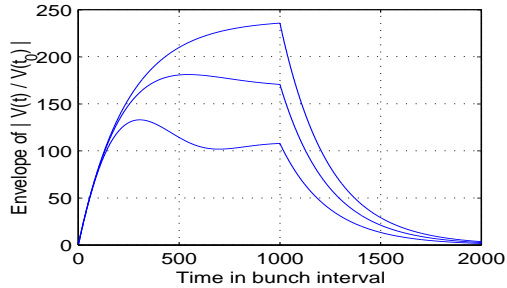

Figure 2. Amplitudes of cavity voltage at resonance (top), and off resonance by $\delta f = f_0/(2Q)$ and by $2\,\delta f$ (bottom).

The asymptotic voltage for infinitely many bunches is shown in Figure 3 for $0.9\omega_{RF} \leq \omega_\lambda \leq 3.2\omega_{RF}$, where $\omega_{RF}$ is the accelerator rf frequency at which the beam is bunched.

$$\lim_{N \to \infty} \left| V_\lambda^N(t) \right| = \left| \exp\left( \frac{\pi}{Q_{L\lambda}} \frac{\omega_\lambda}{\omega_{RF}} - 2\pi i \frac{\omega_\lambda}{\omega_{RF}} \right) - 1 \right|^{-1}.$$

Approximating the asymptotic value around integer values as $\omega_\lambda/\omega_{RF} = n_g + \delta$, we obtain that the output voltage of the resonant mode is enhanced at multiples of the fundamental resonance by the factor of $Q_L/(\pi n_g)$, a large number but decreases with $n_g$, the bunch separation in terms of potential minima. The enhancement of the desired signal, suppression of the unwanted signals, and practical fabrication issues such as tolerance determine the design specification.

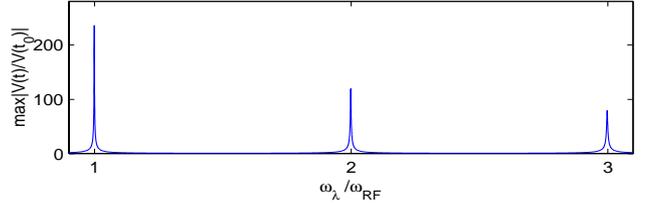

Figure 3. Asymptotic cavity voltage amplitudes.

## 3 QUAD-CAVITY DESIGN

Based on the basic resonance concept described in previous sections, we present a specific design of a Quad-cavity for beam size monitoring or emittance measurement applicable for X-band accelerators operating at 11.424 GHz. First, we choose some design parameters specific to accelerators. The diameter of the beam pipe attached to the Quad-system is chosen as 1cm, large enough for an X-Band linac (with typical iris diameter of 8 mm). Considering resonance enhancement effect and tight tolerance requirement in fabrication, with increasing Q, we choose the external Q value around 750, whose FWHM of the resonance mode energy is 15 MHz. With these chosen parameters, we arrived at an optimized quad-cavity design, shown in Figure 4, that could permit a detection of an rms-beam size under 100 microns, after detailed numerical simulations using electromagnetic field-solvers [4][5]. The main design parameters are summarized in Table 1.

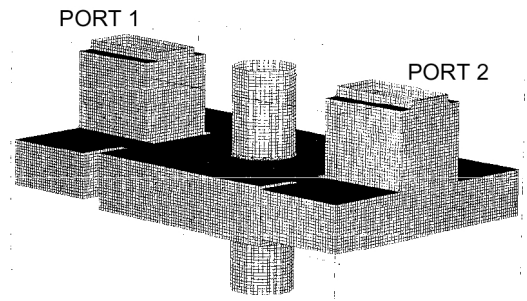

Figure 4. Quad-cavity and two hybrid tee geometry, with shortened waveguides, in mesh from the GdfidL code [4].

The pillbox resonance frequencies of the cavity in Table 1 are 5.8GHz (monopole), 9.1 and 9.3 GHz (dipoles), 11.6 GHz (quad) and 12.84 GHz (next higher mode). The perturbation presented by the four couplings irises, which brings the quad-mode to 11.424 GHz, should not significantly alter this mode spacing.

Table 1. Main parameters of X-band Quad-cavity.

| | |
|---|---|
| Beam pipe diameter = | 1 cm |
| Cavity axial dimension (Lz) = | 0.97 cm |
| Cavity transverse dimensions = | 1.457'' by 1.417'' |
| Waveguide WR62 inner dimensions= | 0.622'' by 0.311'' |
| Quad Resonant Frequency = | 11.424 GHz |

## 4 DESIGN ANALYSIS

Linear accelerators typically have a train of bunches in a pulse. The response of a pulse can be obtained by superposing the port signals of a single bunch at delayed times, each delayed by the bunch intervals. Numerical simulations are performed for a single Gaussian bunch with $\sigma_z = c\sigma_t$ =4mm. Since the GdfidL code used allows transversely only a pencil beam, we offset the beam to x=y=0.5mm, in the cavity coordinates, exciting the quadrupole mode, as well as monopole and dipole modes.

Figure 5 shows the results of various output voltages: output voltage from port 1 (V1) of one bunch (top); V1-V2 of one bunch (next); V1-V2 for 80 nsec pulse (next); and V1+V2 for 80 nsec pulse (bottom). The multi-bunch responses are obtained by superposing the single bunch response at delayed times $1/f_0$=11.424 GHz. The slope of V1-V2 envelope, on a logarithmic scale, gives the external Q of the quad-mode.

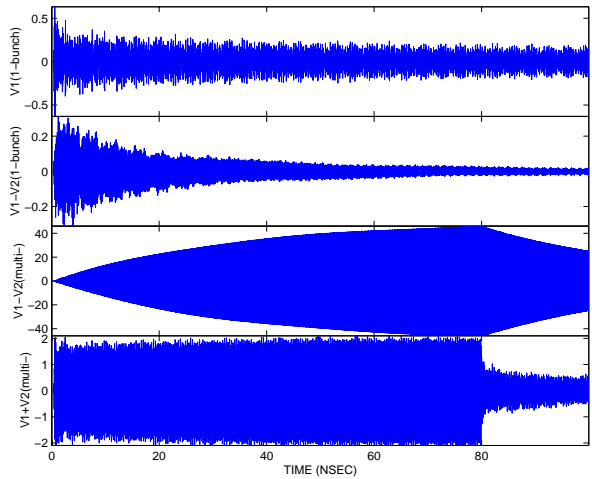

Figure 5. Output voltages: V1 of one bunch (top); V1-V2 of one bunch (next); V1-V2 for 80 nsec pulse (next); and V1+V2 for 80 nsec pulse (bottom).

At resonance, the quadrupole mode signal (V1-V2) increases monotonically with bunches while the dipole component, (V1+V2), reaches a steady value within several bunches, at low values, as seen in Figure 5. The quad-signals show some modulation and saturate at a reduced amplitude when the bunch frequency is mismatched with the cavity resonant frequency as an extension of Figure 2 would predict.

Numerical investigations of various offsets indicate that the level of V1-V2 increases with the offset squared, a quadrupole moment effect, while that of V1+V2 increases with the offset, a dipole moment effect. The third T junction, not modeled, eliminates this dipole signal, passing only the quadrupole signal.

Phase errors in the iris dimensions or locations can be significant. One mil error in one of the iris locations and sizes can modify the resonance frequency by over 10 MHz. On the other hand, phase errors due to inaccuracies in the waveguide lengths allow relatively loose tolerances.

The tolerance for our cavity dimensions is 1 mil for an external Q of 750. The variation of temperature corresponding to 1 mil in the length variation of our Cu-cavity is 40 degree Kelvin.

## 5 DISCUSSION

The rf and mechanical design of the cavity and waveguide network are complete, and a prototype shown in Figure 6 is being fabricated. A tuning device is implemented in the prototype, for fabrication purpose only, to relax tolerance. A device that can pull-and-push, symmetrically, the beampipe attached to the cavity allows the volume of the cavity to be adjusted, thus adjusting the resonance frequency.

The authors thank W. Spence, J. Goldberg, D. Edgell, V. Dolgashev, C. Pearson, M. Ross, and G. Bowden for their input to the project, and W. Bruns for his kind help with the GdfidL code.

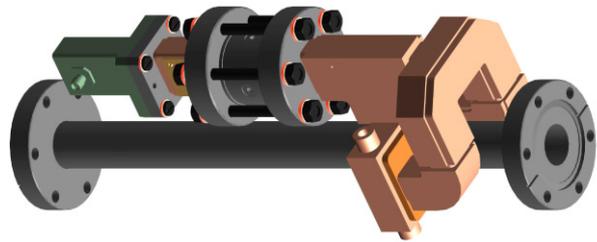

Figure 6. An emittance measurement Quad-cavity system showing rf window assembly and co-ax adaptor.

## 6 REFERENCES

[1] M. Ross, Proc. Adv. Acc. Workshop, AIP Conf. Proc. 279 (AIP, New York, 1992) pp. 807-819.
[2] R. H. Miller, J. E. Clendenin, M. B. James, J. C. Sheppard, Proc. 12th Int. Conf. on High Energy Acc. (Fermilab, Batavia, 1983) SLAC-PUB-3186.
[3] D.H. Whittum, and Y.Kolomensky, Rev. Sci. Instr. 70 (1999) p2300.
[4] W. Bruns, Tech. Univ. Berlin TET-Note 95/14, 1995
[5] HP High Freq. Structure Simulator, V5.4, HP Co.